\documentclass[10pt]{article}
\usepackage{amsfonts}
\usepackage{amsmath}
\usepackage{amssymb}
\usepackage{graphicx}

\def \be {\begin{equation}}
\def \ee {\end{equation}}
\def \bea {\begin{eqnarray}}
\def \eea {\end{eqnarray}}
\def \nn {\nonumber}
\def \la {\langle}
\def \ra {\rangle}
\def \rr {\raise.35ex\hbox{\small $\prime$}\kern-.17em{\mbox{\large $\imath$}}}
\def \del {\partial} 
\def \dels {\partial\kern-.5em / \kern.5em}
\def \As {{A\kern-.5em / \kern.5em}}
\def \Ds {D\kern-.7em / \kern.5em}

\def \a {\alpha}

\def \dag {\dagger}

\def \G {\Gamma}
\def \d {\delta}

\def \m {\mu}
\def \n {\nu}
\def \k {\kappa}
\def \lam {\lambda}

\def \Lap {\nabla^2}
\def \s {\sigma}

\def \om {\omega}

\def \t {\tau}
\def \z {\zeta}

\def \II {I\hspace{-.1em}I\hspace{.1em}}

\def \IIA {\mbox{\II A\hspace{.2em}}}

\def \Area {{\cal A}}  
\def \Length {{\sl l}}

\begin{document}
 
\hfill{arXiv:YYMM.NNNN}

\vspace{20pt}

\begin{center}

{\Large \bf  
Discrete States in
Light-Like Linear Dilaton Background
}
\vspace{30pt}

{\bf Pei-Ming Ho, Sheng-Yu Darren Shih}

\vspace{15pt}
{\em
Department of Physics and Center for Theoretical Sciences, 
National Taiwan University, Taipei 10617, Taiwan, 
R.O.C.
}

\vskip .1in {\small \sffamily pmho@phys.ntu.edu.tw}

\vspace{50pt}
{\bf Abstract}
\end{center}

We study the spectrum of bosonic strings 
in the light-like linear dilaton background
and find discrete states. 
These are physical states which exist only 
at specific values of momentum.
All except one discrete states generate spacetime symmetries. 
The exceptional discrete state 
corresponds to constraints which are 
deformations of conservation laws. 
The constraints resemble those 
arising from symmetries, 
and are equally powerful,
suggesting that our notion of symmetry 
should be generalized.

\setcounter{page}0
\newpage

\section{Introduction}

The original motivation of this work is to understand 
string theory in time-dependent backgrounds. 
Our strategy is to start with the simplest example: 
bosonic string theory in 26 dimensions 
with a light-like linear dilaton background. 
This is an extremely simple case for which 
the worldsheet conformal field theory can be 
found in textbooks \cite{Polchinski}. 
In many aspects, backgrounds depending 
on a light-cone coordinate are much easier than 
generic time-dependent backgrounds. 
Hence there are much recent interest in 
backgrounds with light-like dependence 
\cite{MBB1,MBB,SUGRA,AdSCFT}, 
taking aim at applications to cosmology. 
In particular, the matrix model of type \IIA string theory
in the light-like linear dilaton background 
was proposed in \cite{MBB1}. 
Space-like (e.g. $c=1$ noncritical string \cite{2D}) 
and time-like \cite{timelike} 
linear dilaton backgrounds have also been considered before. 

This work is focused on 
a very simple but intriguing property 
of the light-like linear dilaton background 
that has never been emphasized before 
to the best of our knowledge. 
Recall that in the flat background 
there are 26 spacetime translation symmetries 
generated by 26 physical states 
$\a_{-1}^{\mu}|0, k = 0\rangle$ 
(for an open string or half of a closed string)
corresponding to the vertex operators $\del X^{\mu}$. 
We will refer to these states as ``discrete states'' 
for reasons that will be explained below. 
Remarkably, 
although the translation symmetry 
in a light-cone coordinate is broken
after turning on the dilaton background, 
we still find 26 (physical) discrete states, 
including 25 generators of translation symmetry. 
The additional discrete state which does not 
correspond to a translation symmetry, 
and its implications, will be the focus of this paper. 

In the next section, 
we briefly review discrete states in the 2D string theory
with a space-like linear dilaton background, 
and extend its definition to other backgrounds in string theory. 
These states are physical states 
(i.e., they satisfy Virasoro constraints)
in the old covariant quantization. 
They are not spurious states, 
and their norms can be positive, zero or even negative.
Their existence signals a loop-hole in the no-ghost theorem 
which claims the equivalence between 
the light-cone gauge and old covariant quantization. 
The loop-hole is not fatal, 
and the perturbation theory is still unitary 
due to the discrete nature of these states. 
More importantly, we believe that 
they play a special role in string theory, that is, 
what we see in 2D string theory is not exceptional. 
They correspond to spacetime symmetries 
not only in 2D string theory but also 
in the 26D string theory in flat spacetime. 

In Sec. \ref{discretedilaton}, 
we find discrete states at the massless level also 
in the 26D bosonic string theory 
in the light-like linear dilaton background.
We extend the no-ghost theorem 
for flat spacetime to the linear dilaton background, 
and point out the loop-hole mentioned above. 

The new feature of the dilaton background is that, 
while 25 of the discrete states 
correspond to translation symmetry 
preserved by the background, 
surprisingly an additional discrete state is present, 
without an apparent spacetime symmetry corresponding to it. 
The question is:
is there a less obvious spacetime symmetry 
corresponding to this discrete state, 
or is this discrete state an exception in the special role 
played by all other discrete states? 

To answer this question, 
we compute correlation functions involving this discrete state 
and make the observation in Sec. \ref{ScatteringAmplitudes}
that this state does not decouple from other physical states, 
although it is a zero-norm state.
\footnote{
Here by ``zero-norm state'' we mean a state 
that has vanishing norm. 
Usually, ``zero-norm state'' is a synonym of 
physical spurious states. 
But this discrete state is not spurious.
}
This suggests that the symmetry generated 
by this state is broken, 
if there is really a (hidden) symmetry behind it. 
But we also make the observation that 
there are strong constraints on the kinematics of the theory
\be \label{kchiV}
\sum_a k_a + i \chi_{\cal M} V = 0,
\ee
where $k_a$'s are external momenta, 
$\chi_{\cal M}$ is the Euler character of the string worldsheet 
and $V$ is the gradient of the dilaton field. 
Notice that these constraints, 
although exhibiting the non-conservation of energy-momentum, 
are equally powerful as the $V = 0$ case, 
for which the constraint 
is equivalent to the existence of translation symmetry.

In Sec. \ref{FieldTheory}, 
we elucidate the meaning of these constraints 
in field theory models, 
and propose that this example is calling for 
a generalization of our notion of symmetry. 
Since the constraint (\ref{kchiV}) gives mathematically 
as much information as translation symmetry, 
we should extend our notion of symmetry 
to incorporate deformations of symmetry of this sort. 
Finally, we make concluding remarks in Sec. \ref{Comments}.

\section{Discrete states} \label{discretestates}

The concept of discrete states \cite{DiscreteStates} 
is one of the most crucial ideas in 2D string theory.
The infinitely many discrete states correspond to 
an infinite dimensional symmetry algebra
(the algebra of area-preserving diffeomorphisms) \cite{walgebra},
which can be used to uniquely determine
all correlation functions in 2D string theory.
In this paper, we would like to generalize 
the notion of discrete states to 26D bosonic string theory.

The discrete states in 2D string theory have a few salient features:

\begin{enumerate}
\item
In the formulation of the old covariant quantization, 
the physical spectrum
admits additional physical states
at certain special (discrete) values of the momentum
due to a degeneracy of Virasoro constraints 
or spurious states,
in contrast with generic values of momentum.

\item
These states indicate a breakdown of the equivalence 
between the old covariant quantization 
(or the BRST quantization) and the light-cone gauge,
which is essentially the no-ghost theorem.

\item
They are associated with symmetries of the theory.
\end{enumerate}

In the light-cone gauge, 
zero-norm and negative-norm states which are
potential threats to the unitarity
are eliminated by gauge-fixing.
This means that, in the old covariant quantization,
physical conditions (Virasoro constraints) are strong enough, 
and gauge transformations (spurious states) are plenty enough 
to eliminate all the negative-norm states.
As mentioned above in the first property of discrete states,
the existence of the discrete states is 
either due to a weakness of the Virasoro constraint or 
the lack of spurious states at particular momenta.
The second property is therefore closely related to the first.
The third property is less directly related to the other two,
but is the main reason why discrete states 
plays a very important role in 2D string theory. 
In the following, we will extend the use of this terminology 
and refer to a state 
in any string theory as ``discrete state''
if the first two properties are satisfied.

\subsection{Discrete states in 2D with space-like linear dilaton background}
\label{review}

For completeness, we briefly review the idea of 
discrete states in 2D string theory.

Consider the worldsheet action for a bosonic string 
in the linear dilaton background
\footnote{
For noncritical strings, such as 2D strings,
one needs to add the Liouville potential $\mu e^{\a X}$ 
to the Lagrangian, 
but in those cases we will consider the spectrum 
of strings in the region far away from the Liouville wall,
where the Liouville potential can be ignored.
Hence the potential is irrelevant to our consideration.
}
\be \label{action}
S=\frac1{4\pi\a'}\int_{\cal M}d^2\s\,\sqrt{g} 
(g^{ab}\del_a X\cdot \del_b X+\a'R(\s)V\cdot X)
+\frac1{2\pi}\oint_{\del \cal M}ds\, \k(\s)V\cdot X,
\ee
where 
$R(\s)$ is the worldsheet curvature and 
$\k(\s)$, the geodesic curvature of the boundary.
This is a conformal field theory (CFT) with energy-momentum tensor 
\be \label{T}
T(z)=-\frac1{\a'}:\del X^\m \del X_\m :+V_\m \del^2 X^\m,
\ee
and central charge
\be 
c=d+6\a'V_\m V^\m.
\ee
Since the linear dilaton background leaves the ghost action intact, 
the ghost central charge remains to be $-26$. 
Demanding that the total central charge 
vanishes to give an anomaly-free theory, 
we have the following condition on space-time dimension
\be
d=26-6\a'V\cdot V.
\label{criticaldim}
\ee
In particular, if $V=(0, \frac2{\sqrt{\a'}} )$ , 
we obtain a 2D string theory 
in space-like linear dilaton background. 
It is easy to see in the light-cone gauge that
there can be no physical polarizations in 2D 
for generic momentum.
For example, consider the first excited state
$\z\cdot\del Xe^{ik\cdot X}$ for an open string
(normal ordering is always assumed).
The Virasoro conditions read
\begin{align}
(L_0-1)|\mbox{phys}\ra=0 \;&\Rightarrow k\cdot (k+iV)=0,\label{1} \\
L_1|\mbox{phys}\ra=0 \;&\Rightarrow \z\cdot(k+iV)=0,\label{2}
\end{align}
and a gauge transformation is implied by a spurious state
\begin{align}
L_{-1}|\psi\ra=\mbox{null}\;&\Rightarrow \z^\m \sim \z^\m+\lam k^\m.
\label{3}
\end{align}
One can see that for a generic $2$-vector $k^\m$, 
the constraints above have no solutions for the polarization vector $\z$.
However, 
when $k^\m=-iV^\m$, 
the physical conditions (\ref{1}) is trivially satisfied.
On the other hand, when $k^\m=0$, 
the spurious state (\ref{3}) does not exist.
Thus we obtain two discrete states at 
the specific momenta 
$k^\m=-iV^\m$ and $k^\m=0$. 
A short analysis shows that in these cases 
one can choose the {\em material gauge} 
$\z^1=0$. 
In fact, it can be shown that 
the material gauge works at all massive levels. 
All we need to do in order to find discrete states is looking for
primary fields in the Fock space of $X^0$. 
It turns out that there are infinitely many discrete states 
corresponding to an infinite dimensional symmetry of 2D strings.

\subsection{Discrete states in 26D flat space}
\label{translsymm}

It may appear to some readers that 
the existence of the discrete states rely on 
the choice of a very special background.
But we will show here that there are discrete states 
even for the background of flat spacetime.

For the 26D flat background,
a generic state at the massless level ($k^2 = 0$) 
is of the form
\be
|\psi\ra=
\zeta \cdot\a_{-1}
|0;k\ra.
\ee 
For zero momentum $k = 0$, 
the Virasoro constraint 
\begin{align}
L_1|\psi\ra = 0
\end{align}
is satisfied for any polarization $\zeta^\m, \,(\m=0,1,\cdots25)$. 
Furthermore, there is no 
spurious state with zero momentum, 
since 
$$
L_{-1}|0; 0\ra = 0.
$$
Therefore, we get an enlarged 
spectrum of physical states at $k=0$:
there are 26, instead of 24, 
physical polarizations of the massless vector field at $k = 0$. 
We will refer to these states as discrete states, 
because they share the same features as the discrete states in 2D.

This may seem a bit weird, 
since we have the no-ghost theorem asserting that 
the spectrum of old covariant quantization
is isomorphic to the spectrum 
in the light-cone gauge,
which has only 24 physical polarizations.
How can there be a mismatch at discrete states? 
Apparently, there is a crack in the proof of the no-ghost theorem. 
We will show where the crack is in Sec. \ref{noghost}.
More importantly, is unitarity lost due to discrete states?
All the discrete states with space-like polarizations have positive norm,
but the one with a time-like polarization has negative norm.
(One can also superpose them to obtain a state with 
a light-like polarization, 
which has zero norm.)
Hence there is a ``ghost'' in the spectrum.
Unitarity would be violated if
the ghost can be generated in a scattering process.
Luckily, the probability of generating a discrete state is zero, 
because the phase space available for the discrete state 
has measure zero. 
Therefore the existence of discrete states does not 
imply the violation of unitarity, 
regardless of whether they have positive or negative norms.

If discrete states cannot be generated in scattering, 
does this mean that they are meaningless in string theory?
Like discrete states in 2D, 
discrete states in 26D flat spacetime 
are also generators of spacetime symmetry.
The vertex operators corresponding to 
the discrete states $\a^{\mu}_{-1}|0;0\ra$ are $\del X^{\mu}$, 
which are the conjugate momenta of $X^{\mu}$.
Thus the discrete states at $k = 0$ have a clear physical meaning: 
they generate the spacetime translation symmetry.
All the 26 discrete states must therefore decouple 
from all physical states, 
as a statement of momentum conservation.


\section{Discrete states in light-like linear dilaton background}
\label{discretedilaton}

In the previous section we learned two things.
First, discrete states exist not only in 2D,
but also in 26D.
Second, discrete states seem to always generate symmetries.
In the following we will test these two observations 
in the light-like linear dilaton background.

The worldsheet action of a bosonic string
in the linear dilaton background was given in (\ref{action}).
From (\ref{criticaldim}), 
it is easy to see that 
if we have a light-like linear dilaton background 
($V\cdot V=0$), 
we get a 26D critical string theory. 
Let us review the basic ingredients of this CFT.
Taking the Laurent expansion of (\ref{T}),
\be
T(z) =\sum^{\infty}_{m=-\infty}{L_m \over z^{m+2}}, \qquad
L_m=\oint dz\, z^{m+1}T(z),
\ee
one obtains the Virasoro generators
\be
L_m=\frac12\sum^{\infty}_{n=-\infty}\, 
\a_{m-n}^\m\a_{\m n}+i\sqrt{\frac{\a'}2}(m+1)\,V^\m \a_{\m m}.
\ee 

The OPE of $X$, $T$, and $L_n$
are independent of the linear dilaton background,
because the linear dilaton background is a topological effect.
It does not change the field equations and
we have the same mode expansion
\be
\del X^\m(z)=-i\left(\frac{\a'}2\right)^{1/2}
\sum^{\infty}_{m=-\infty}{\a_m^\m \over z^{m+1}},
\ee
and the same canonical commutation relations
\be
[\a_m^\m,\a_n^\n]=m\d_{m+n}\eta^{\m\n} 
\ee
as in flat spacetime.
(Here $ p^\m=\left(\frac2{\a'}\right)^{1/2}\a_0^\m$.)
One can see this also by separating 
$X^\m(\s)$ into the homogeneous solution $X_h^\m(\s)$
and the special solution $X_s^\m(\s)$ (see the appendix). 
$X_h^\m(\s)$ has the same field equations and boundary conditions
as in flat spacetime,
and hence the same OPE. 
Using conformal transformations, 
we can set $X_s^\m(\s)=0$ everywhere 
except a point with curvature singularity,
which can be pushed to infinity such that 
$X_s^\m(\s)$ does not affect the OPE.

\subsection{Discrete states at massless level}

To find discrete states in this theory,
we take a closer look at the physical conditions 
and spurious states for the light-like linear dilaton background. 
The Virasoro constraints are
$$
(L_n-\d_{n,0})|\mbox{phys}\ra=0, \quad \forall n\geq 0.
$$      
We will use the convention that
$\a'=\frac12$ for the open string theory.

The spurious states are those orthogonal to all physical states.
In flat spacetime,
the spurious states are $\{L_{-m}|\chi\ra\}$, 
where $m>0$ and $|\chi\ra$ is an arbitrary state.
Since 
$L_m^{\dag} = L_{-m}$, 
the inner product of a spurious state with any physical state $|\psi\ra$ is 
\be
\la \psi|L_{-m}|\chi\ra = (L_m|\psi\ra)^{\dag}|\chi\ra = 0,
\ee
due to Virasoro constraints on the physical state.
However,
in the linear dilaton background, 
the adjoint of $L_{-m}$ equals $L_m$ with shifted momentum 
$\hat p^\m \rightarrow \hat p^\m-iV^\m$.
That is, since $\a_{\mu m}^{\dagger}=\a_{\mu (-m)}$, 
we have
\begin{align} 
[L_{-m}(\hat{p})]^{\dag} &=
L_m(\hat{p})-2i\sqrt{\frac{\a'}2}V^\m\a_{\m m} \nn\\
&=(\hat p^\m-iV^\m)\a_{\m m}+
\frac12 \sum_{n\neq 0}\a_{m-n}\cdot\a_n + 
i\sqrt{\frac{\a'}2}(m+1)\,V^\m \a_{\m m} \nn \\
&= L_m(\hat{p}-iV).
\end{align}
The two-point function of vacuum states
in the light-like linear dilaton background is given by
\be \label{innerp}
\la 0;k|0;k'\ra=(2\pi)^d \d(k'-k+iV),
\ee
where $\la 0;k|$ and $|0;k'\ra$ are left and right eigenstates of 
the momentum operator $\hat{p}$
\be
\la 0;k|\hat{p} = \la 0;k|k, \qquad
\hat{p}|0;k' \ra = k'|0;k'\ra.
\ee
Since (\ref{innerp}) should be interpreted as 
the norm squared $\left| |0;k'\ra \right|^2$ 
when it is non-vanishing,
the Hermitian conjugate of a state 
should have its momentum shifted by $-iV^\m$, i.e.
$$
(|0;k\ra)^\dag=\la0;k-iV|, \qquad \mbox{or equivalently,} 
\qquad \la0;k|=(|0;k+iV\ra)^\dag.
$$
This cancels the momentum shift of $L_{-m}^\dag$exactly. 
Indeed
\be 
\la \psi;k|L_{-m}(\hat p)|\chi; k'\ra =
\la \psi;k|[L_m(\hat p -iV)]^{\dag}|\chi; k'\ra 
=\left\{L_m(\hat p -iV)|\psi; k+iV\ra\right\}^\dag |\chi; k'\ra = 0,
\ee
by the physical condition on $|0;k\ra$.
The spurious states are thus still of the same form 
$\{ L_{-m}|\chi\ra \}$ as in flat spacetime.

Now we study the physical spectrum at the massless level. 
A generic first excited state $|\psi\ra$ 
can be written as $\z\cdot\a_{-1}|0;k\ra$,
which is subject to the physical conditions
\begin{align}
(L_0-1)|\psi\ra=0 &\Rightarrow k\cdot(k+iV)=0, \\
L_1|\psi\ra=0 &\Rightarrow \z\cdot(k+iV)=0. 
\end{align}
The only spurious state at this level is
\be
L_{-1}|\psi\ra= k\cdot\a_{-1}|\psi\ra 
\Rightarrow \z^\m\sim\z^\m+\lam k^\m.
\ee
This is formally the same as the 2D case,
but now we have 26 space-time dimensions, 
and the above equation does have non-trivial solutions 
for generic momentum $k$. 
The constraint $\z\cdot(k+iV)=0$ reduces one degree of freedom 
and the gauge symmetry $\z^\m\sim\z^\m+\lam k^\m$ 
eliminates another.
As a result, they admit 24 physical polarizations, 
in agreement with the light-cone gauge.

However, at $k=-iV$ the constraint on polarization 
becomes trivial 
and we have one additional physical excitation
\be \label{D-}
|D^-\ra \equiv \a_{-1}^- | 0; -iV \ra \quad \leftrightarrow \quad
\del X^-e^{i(-iV)\cdot X}=\del X^-e^{-V^-X^+}.
\ee
Here we choose the convention that 
the only nonvanishing component of $V$ is $V^-$.
Similarly, at $k=0$ there is no spurious state, 
and the physical spectrum is enlarged.
The physical states with $k = 0$ are
\begin{align}
|D^+\ra \equiv \a_{-1}^+ | 0; 0 \ra \quad &\leftrightarrow \quad
\del X^+, \\
|D^i\ra \equiv \a_{-1}^i | 0; 0 \ra \quad &\leftrightarrow \quad
\del X^i,
\end{align}
where the index $i$ is used for the 24 transverse directions. 
We will refer to these states as discrete states 
for the same reason we used this terminology 
for the flat spacetime in Sec. \ref{translsymm}. 
All these discrete states are one-to-one matched 
with those in 26D flat spacetime.
Obviously $|D^+\ra$ and $|D^i\ra$ are still
generators of translation symmetries in spacetime. 
It is unexpected that $|D^-\ra$ is present 
since the translation symmetry in $X^+$ 
is broken by the dilaton background.



\subsection{No-ghost theorem}\label{noghost}

Let us now examine the argument of 
no-ghost theorem \cite{GoddardThorn} 
following the presentation in \cite{Polchinski},
and see how the equivalence between light-cone gauge 
and BRST quantization breaks down on discrete states.
We will skip the proof of the equivalence between
the old covariant quantization 
and BRST quantization,
which does not suffer the same problem. 

First, we try to prove the no-ghost theorem 
for the linear dilaton background,
by adapting the proof for 26D flat space in \cite{Polchinski}.
The proof is composed of two parts.
The first part of the proof is to find 
the cohomology of a simplified BRST operator $Q_1$, 
which has the same physical content as the light-cone gauge theory. 
The second is to show that 
the cohomology of the full BRST charge $Q_B$ 
is identical to that of $Q_1$. 

To begin, define the light-cone oscillators
\be
\a_m^\pm=2^{-1/2}(\a_m^0\pm\a_m^1).
\ee
They satisfy the commutation relation
\be
[\a_m^+,\a_n^-]=-m\d_{m+n}, \qquad
[\a_m^+,\a_n^+]=[\a_m^-,\a_n^-]=0.
\ee
We also define the number operator
\be
N^{lc}=\sum^{\infty}_{\stackrel{m=
-\infty}{m\neq0}}\frac1m\a_{-m}^+\a_m^- .
\ee
It counts the number of $X^-$ excitations 
minus the number of $X^+$ excitations.

Now decompose the BRST generator according to 
the value of $N^{lc}$ as
\be
Q_B=Q_1+Q_0+Q_{-1},
\ee
where $Q_j$ changes $N^{lc}$ by $j$ units
\be 
[N^{lc}, Q_j]=jQ_j.
\ee
Expanding $Q_B^2=0$ gives
\be
\left(Q_1^2\right)+\left(\{Q_1,Q_0\}\right)+
\left(\{Q_1,Q_{-1}\}+Q_0^2\right)+\left(\{Q_0,Q_{-1}\}\right)+
\left(Q_{-1}^2\right)=0.
\ee
Each group in parentheses has a different $N^{lc}$ number 
and must vanish separately. 
In particular, $Q_1$ and $Q_{-1}$ are nilpotent, 
hence each defines a cohomology of its own. 
They are given by
\begin{align}
Q_1&=-\sqrt{2\a'}k^+
\sum^{\infty}_{\stackrel{m=-\infty}{m\neq0}}\a_{-m}^-c_m, 
\nn\\
Q_{-1}&=-\sqrt{2\a'}\sum^{\infty}_{\stackrel{m=-\infty}{m\neq0}}
\left[k^-+\frac i2V^-(1-m)\right]\a_{-m}^+c_m.
\end{align}

Assuming that $k^+\neq0$, we introduce the operator
\be \label{Rdef}
R={1\over(2\a')^{1/2}k^+}\sum^{\infty}_{\stackrel{m=-\infty}{m\neq0}}
\a_{-m}^+b_m
\ee
such that
\begin{align}
S\equiv\{Q_1,R\}&=\sum^{\infty}_{m=1}
(mb_{-m}c_m+mc_{-m}b_m-\a_{-m}^+\a_m^--\a_{-m}^-\a_m^+)\nn\\
&=\sum^{\infty}_{m=1}m(N_{bm}+N_{cm}+N_m^++N_m^-).
\end{align} 
The normal ordering constant is determined by noting that 
$Q_1$ and $R$ both annihilate the ground state. 
Because $Q_1$ commutes with $S$, 
we can calculate the $Q_1$ cohomology 
within each eigen-space of $S$, 
and the full cohomology is the union of the result. 

If $|\psi\ra$ is $Q_1$-closed with $S|\psi\ra=s|\psi\ra$, 
then for nonzero $s$
\be
|\psi\ra=\frac1s\{Q_1,R\}|\psi\ra=\frac1sQ_1R|\psi\ra,
\ee
and so $|\psi\ra$ is actually $Q_1$-exact. 
Therefore, the $Q_1$ cohomology can be 
nontrivial only at $s=0$. 
Clearly, the operator $Q_1$ annihilates all states 
in $\mbox{\em ker}\,(S)$, 
so they are all $Q_1$-closed and 
there are no $Q_1$-exact states in this space. 
Therefore, we have 
\be
\mbox{\em cohomology}\,(Q_1)\cong \mbox{\em ker}\,(S).
\ee
On the other hand, by the definition of $S$, 
the $s=0$ states have no longitudinal or ghost excitations 
--- 
$\mbox{\em ker}\,(S)$ is just the Hilbert space 
of the light-cone gauge ${\cal H}_{lc}$.  
Therefore the $Q_1$-cohomology is ${\cal H}_{lc}$. 
This proves the no-ghost theorem for $Q_1$. 

To complete the proof, 
we have to show that the $Q_1$-cohomology is 
isomorphic to the $Q_B$-cohomology. 
To achieve this goal, let's consider the operator
\be
S+U\equiv\{Q_B,R\},
\ee
where $U=\{Q_0+Q_{-1},R\}$ and 
it lowers $N^{lc}$ by one or two units. 

For each state $|\psi_0\ra$ 
in $\mbox{\em ker}\,(S)$, 
\footnote{
The factor $S^{-1}$ makes sense because 
it always acts on states with $N^{lc}<0$.
} 
one can construct another state 
\be
|\psi\ra=(1-S^{-1}U+S^{-1}US^{-1}U-\cdots)|\psi_0\ra,
\ee
which is annihilated by $S+U$, 
i.e. $|\psi\ra\in \mbox{\em ker}\,(S+U)$. 
Now, consider a $Q_B$-closed state $|\psi\ra$.
Following the same arguments as above, 
with $S+U$ replacing $S$, 
one sees that 
\be
\mbox{\em cohomology}\,(Q_B) \cong
\mbox{\em ker}\,(S+U) \cong
\mbox{\em ker}\,(S) \cong
\mbox{\em cohomology}\,(Q_1) \cong 
{\cal H}_{lc},
\ee
and the proof is completed. 
But all this was based on the assumption 
that $k^+ \neq 0$ so that (\ref{Rdef}) is well defined.

Incidentally, we remark that 
for the flat spacetime, the discrete states with $k = 0$ 
can never have $k^+ \neq 0$ in any Lorentz frame, 
and thus the proof breaks down. 

Since the linear dilaton background breaks the Lorentz symmetry,
it might happen that some states with momenta $k$ has
$k^+=0$ but we cannot use Lorentz transformation 
to make $k^+$ nonzero even when $k$ is not identically zero. 
In such cases the role of $Q_1$ can be replaced by $Q_{-1}$, 
with the corresponding operator $R$ defined as
\be
R=\frac1{\sqrt{2\a'}}\sum_{m\neq0}{\a_{-m}^-b_m\over k^-
+\frac i2V^-(1+m)},
\ee
so that 
\be
S\equiv\{Q_{-1},R\}=
\sum^{\infty}_{m=1}m(N_{nm}+N_{cm}+N_m^++N_m^-).
\ee
The same argument above works
as long as 
$k^-+\frac i2(m+1)V^-\neq0$ 
for all nonzero integer $m$.

On the other hand, if
\be
k^+=0, \qquad 
k^-=-\frac i2(1+m)V^-,
\ee
and all other components of $k$ vanish,
the no-ghost theorem breaks down. 
At the massless level, only oscillation operators with
indices $|m|\leq1$ matters.
We discuss each case separately:
\begin{enumerate}
\item
$m=-1\;\Rightarrow k^\m=0$. 
This corresponds to the operators $\del X^\m$, 
corresponding to the states $|D^i\ra$ and $|D^+\ra$. 
In flat space, these states are just the generating currents 
of translation symmetry we discussed in Sec. \ref{translsymm}. 

\item
$m=1\;\Rightarrow k^\m=-iV^\m$. 
This is just the discrete state $|D^-\ra$ (\ref{D-}).    
\end{enumerate}

In summary, 
at the massless level, 
the equivalence between BRST quantization and light-cone gauge 
breaks down exactly at discrete states.

\section{Scattering amplitude in light-like linear dilaton background}
\label{ScatteringAmplitudes}

In the previous section, 
we found that there are 26 discrete states
in the light-like linear dilaton background.
While 25 of them correspond to spacetime translation symmetry,
one is tempted to make the conjecture that 
the state $|D^-\ra$ is also a generator of a certain symmetry.
If this is indeed the case, 
all states of the theory can be organized according to
the representations (charge) of the symmetry generated by $|D^-\ra$. 
We do not know yet what is the symmetry transformation, 
but in principle such information can be deduced from 
the knowledge of correlation functions. 
For instance, the discrete states should decouple 
from all states in the trivial representation. 
In this section, 
we will try to investigate properties of the group 
by studying correlation functions. 

For ordinary vertex operators 
(those which are not discrete states), 
their correlation functions in the light-like linear dilaton background
and those in the flat spacetime are related to each other 
by a simple formula
\cite{MBB}
\be \label{SC}
S^{g, n}_{\mbox{\tiny dilaton}} = 
S^{g, n}_{\mbox{\tiny flat}} \cdot C, \qquad
C = 
\int_{-\infty}^{\infty} d\t_* \;
e^{-(2g-2+n)V\cdot p \t_*},
\ee
where $S^{g,n}$ is the diagram of genus $g$ 
for an $n$-point function. 
This relation can be easily seen from a calculation 
in the light-cone gauge $X^+ = p^+ \t$.
In this gauge, the remaining worldsheet dynamical fields $X^i$
are insensitive to the linear dilaton background.
The only effect of the dilaton background to the computation 
of a scattering amplitude
is the appearance of an exponential factor $e^{-V\cdot p \t}$
at each point $\t$ of the insertion of a joining/splitting operator
in a light-cone gauge diagram.
A diagram of genus $g$ for an $n$-string scattering process
involves the integrals of $(2g-2+n)$ parameters 
for the insertion of joining/splitting operators. 
The overall effect is to multiply the flat space amplitude 
by the factor $C$ (\ref{SC}), 
where $\t_*$ is the average of insertion points.
However, this is only a formal relation since 
the on-shell conditions are modified 
by the linear dilaton background, 
and the scattering amplitude is only defined on-shell.
Furthermore, as we have seen in the previous section, 
the discrete states are missing in the light-cone gauge.
The correlation functions of interest to us need to 
be calculated using the old covariant quantization.

In the appendix, we develop the path integral formulation 
for the light-like linear dilaton background.
The most salient feature of this background is that 
the energy-momentum conservation law is modified 
\be \label{mom-conserv}
{\sum}_a k_a^\m+i{\chi}_{\cal M}V^\m = 0,
\ee
where $\chi_{\cal M}$ is the worldsheet Euler character.
For a Rieman surface with $g$ genus, $b$ boundaries 
and $c$ cross-caps, 
the Euler character is 
\be
\label{Euler}
\chi_{\cal M} = 2 - 2g - b - c.
\ee
We will refer to this constraint (\ref{mom-conserv}) 
as the {\em (non-)conservation law} of momentum. 
It is valid for both open and closed strings. 

The (non-)conservation law 
can be easily derived by integrating out the zero modes 
of $X^{\mu}$ in the path integral.
Apparently, there is a one-to-one correspondence
between the 26 components of (\ref{mom-conserv}) 
and the 26 discrete states.

In the following we will compute open string amplitudes 
at tree level.
The worldsheet is a disk $D_2$ and $\chi_{D_2} = 1$.
We will only consider correlation functions composed of 
tachyons and massless vectors. 
The time reversal symmetry on the worldsheet 
implies that a correlation function of 
an arbitrary number of tachyons 
and an odd number of massless vectors vanishes identically.

\subsection{Three-point functions}



As a warmup, we start with the simplest case -- 
the three tachyon scattering amplitude. 
Up to a delta function imposing the
momentum (non-)conservation law (\ref{mom-conserv}), 
it should be a constant that defines 
the coupling strength for the 3-tachyon interaction.
The path integral gives
\begin{align}
S_{D_2}(k_1;k_2;k_3)
&=\left\la:c(y_1)e^{ik_1\cdot X(y_1)}::c(y_2)e^{ik_2\cdot X(y_2)}:
:c(y_3)e^{ik_3\cdot X(y_3)}:\right\ra
+ \mbox{permutation} \nn\\
&=2ig_o^3C_{D_2}(2\pi)^d\d^d\left(\sum^{3}_{a=1}k_a^\m
+ i{\chi}_{D_2}V^\m\right)
\,|y_{12}|^{2\a'k_1\cdot k_2+1}\,|y_{23}|^{2\a'k_2\cdot k_3+1}
\,|y_{31}|^{2\a'k_3\cdot k_1+1}.
\end{align}                   
The on-shell condition for tachyons asserts that
\be
\frac1{\a'}=k\cdot(k+iV)=(k+\frac i2V)^2.
\ee
Using the momentum (non-)conservation law, 
we can rewrite the quantities $k_a\cdot k_b$, e.g.
\begin{align}
2\a'k_1\cdot k_2
=-1+2i\a'({\chi}_{D_2}-1)V\cdot k_3 = -1
\end{align}           
for the disk diagram.
The three tachyon amplitude is thus
\be
ig_0^3C_{D_2}(2\pi)^d\d^d
\left(\sum^{3}_{a=1}k_a^\m+iV^\m\right).
\ee 


Similarly, the correlation function of one tachyon 
with two massless vectors is 
\be
S_{D_2}(k_1; \xi, k_2; \zeta, k_3) 
\sim (-\xi\cdot k_1 \zeta\cdot k_1 + \xi\cdot \zeta).
\ee


\subsection{Four-point functions}

Before computing four-point functions, 
we introduce Mandelstam variables as follows
\begin{align}
s&=\a'(k_3+k_4)\cdot(k_1+k_2)+1, \label{M1} \\ 
t&=\a'(k_2+k_4)\cdot(k_1+k_3)+1, \label{M2} \\
u&=\a'(k_1+k_4)\cdot(k_2+k_3)+1. \label{M3}
\end{align}

Note that due to the modification of momentum conservation, 
the generalization of the original Mandelstam variables 
to the linear dilaton background is ambiguous.
In flat space, it is equivalent to write 
$s = - \a' (k_1+k_2)^2$ 
or $s = - \a' (k_3+k_4)^2$. 
But since we have $k_1 + k_2 + k_3 + k_4 = -iV$
in the linear dilaton background, 
these two expressions are not the same. 

We find the definitions above convenient.
They transform simply under
the exchange of two momenta. 
For example, 
under the exchange $(k_1\leftrightarrow k_2)$, 
they transform as 
\be
s\leftrightarrow s, \qquad 
t\leftrightarrow u, \qquad 
u\leftrightarrow t. \nn
\ee


\subsubsection{Two tachyons and two massless vectors}


The correlation function of 2 tachyons and 2 massless vectors is
\begin{align}
S_{D_2}^4&=g_o^4e^{-\lam}\int^{\infty}_{-\infty}dx
\left\la:c(0)e^{ik_1\cdot X(0)}::c(1)e^{ik_2\cdot X(1)}::c(\infty)
\,\xi\cdot\del Xe^{ik_3\cdot X(\infty)}::\z\cdot
\del X(x)e^{-ik_4\cdot X(x)}:\right\ra \nn\\
&\;\;\;+\mbox{permutation} \nn\\
&= K(2\pi)^d\d^d\left(\sum^{4}_{a=1}k_a^\m+iV^\m\right)
\left(-4{\a'}^2\left[\z\cdot k_1\xi\cdot k_2\,st
+\z\cdot k_2\xi\cdot k_1\,su+\z\cdot k_3\xi\cdot k_4\,tu\right]
+2\a'\z\cdot\xi\,tu\right) \nn\\
&\;\;\;\times\left[{\G(-s)\G(-t)\over\G(u+1)}
+{\G(-s)\G(-u)\over\G(t+1)}-{\G(-u)\G(-t)\over\G(s+1)}\right]
+\mbox{permutation}, 
\end{align}
where $K = ig_o^4C_{D_2}$.
To derive this formula, we used the relation 
\be
s+t+u=\a'\sum_im_i^2+3
=1 \label{SumM} 
\ee 
among the Mandelstam variables. 
According to (\ref{M1}-\ref{M3}) and (\ref{SumM}), they are
\bea
s=&-2\a'k_3\cdot k_4+1&=-2\a'k_1\cdot k_2 -1, \nn\\
t=&- 2\a'k_2\cdot k_4&=-2\a'k_1\cdot k_3, \nn\\
u=&- 2\a'k_1\cdot k_4&=-2\a'k_2\cdot k_3.
\eea 
The permutation in the expression above 
can be carried out in two ways.
The first way is to exchange the positions of two tachyons, 
which can be done by 
simply exchanging $k_1\leftrightarrow k_2$ 
and $t\leftrightarrow u$, 
and the net result is to simply duplicate the above formula. 
On the other hand, we can choose to exchange the positions 
of one tachyon and one photon.
We checked explicitly that 
these two different prescriptions of permutation
give the same final answer, 
and this also serves as a check of the conformal symmetry of 
the light-like linear dilaton background.


Next, we want to check the gauge symmetry, 
which corresponds to the decoupling of null states. 
This is not a trivial check in that 
we have modified the definition of bras and kets 
in order to make sense of null states. 
We separate a polarization $\z^\m$ into the transverse and longitudinal parts
\be
\z=\z^\bot+\z^\parallel k_4,
\ee
where $\z^\bot$ denotes the transverse polarization vector,
and $\z^\parallel$ parametrizes the magnitude of 
the longitudinal part. 
Similarly we have $\xi=\xi^\bot+\xi^\parallel k_3$. 
Under this separation, 
we can simplify the formulas using the relation
\begin{align}
\z\cdot k_3=\z^\bot\cdot k_3+\z^\parallel k_4\cdot k_3 
=\z^\bot\cdot k_3+\frac{\z^\parallel}{2\a'}(1-s).
\end{align} 

Now, the amplitude reads
\begin{align}
S_{D_2}^4\sim&-4\a'^2(\z^\bot\cdot k_1\xi^\bot\cdot k_2)
\,st+2\a'(\z^\bot\cdot k_1\xi^\parallel+\xi^\bot\cdot k_2\z^\parallel)
\,stu-\z^\parallel\xi^\parallel\,u^2st \nn\\
&-4\a'^2(\z^\bot\cdot k_2\xi^\bot\cdot k_1)\,su
+2\a'(\z^\bot\cdot k_2\xi^\parallel+\xi^\bot\cdot k_1\z^\parallel) 
\,stu-\z^\parallel\xi^\parallel\,t^2su \nn\\
&-4\a'^2(\z^\bot\cdot k_3\xi^\bot\cdot k_4)
\,ut+2\a'(\z^\bot\cdot k_3\xi^\parallel+\xi^\bot\cdot k_4\z^\parallel) 
\,stu-\z^\parallel\xi^\parallel\,s(s-1)tu +2\a'\z^\bot\cdot \xi^\bot\, ut. 
\end{align}
The $\xi^\parallel$ terms vanish due to 
the momentum (non-)conservation law (\ref{mom-conserv})
and the physical condition 
\be
\z^\bot\cdot(k_1+k_2+k_3)=-\z^\bot\cdot(k_4+iV)=0.
\ee
Similarly the $\z^\parallel$ terms also vanish. 
The $\z^\parallel\xi^\parallel$ term vanishes because  
\be
u^2st+t^2su+s(s-1)tu=stu[(s+t+u)-1]=0.
\ee

Therefore, the null state $k\cdot\a_{-1}|0;k\ra$ 
indeed decouples from the physical spectrum. 
The scattering amplitude is thus
\be \label{2T2V}
S_{D_2}^4\sim -4\a'^2(\z^\bot\cdot k_1\xi^\bot\cdot k_2)\,st
-4\a'^2(\z^\bot\cdot k_2\xi^\bot\cdot k_1)\,su
-4\a'^2(\z^\bot\cdot k_3\xi^\bot\cdot k_4)\,ut
+2\a'\z^\bot\cdot \xi^\bot\,ut. 
\ee
This expression can be applied to several special cases 
of interest to us. 

If $k_4 = 0$, i.e., if the 4-th particle 
is one of the 25 discrete states $(|D^+\ra, |D^i\ra)$,
we find $t = u = 0$, 
and the amplitude vanishes identically,
as a reflection of the translation symmetry in spacetime.
On the other hand, 
if $k_4 = -iV$ and $\zeta = - \hat{e}^+$, i.e., 
if the last particle is the discrete state $|D^-\ra$, 
the amplitude is not identically zero.
To check this, it suffices to give an example 
with $S_{D_2}^4\neq 0$.
Consider the case when
\be \label{k1k2}
k_1 = i \lam W - i(1/2-\mu) V + {\bf k}, \qquad
k_2 = -i \lam W - i(1/2+\mu) V - {\bf k},
\ee
where $W = -\hat{e}^- /V^-$, 
and ${\bf k}$ is a vector perpendicular to both $W$ and $V$.
Since $k_4 = -iV$, the modified momentum conservation 
implies that $k_3 = -(k_1+k_2) = iV$ and hence $s=1$.
The ansatz above for the momenta makes sure that 
the on-shell conditions for all momenta $k_a$ are satisfied 
as long as 
\be
\mu = \frac{1}{2\lam}\left({\bf k}^2 - \frac{1}{\a'}\right).
\ee
The $\xi^-$ component of the polarization $\xi$ 
of the 3rd particle is a pure gauge, 
and for the momenta assignment (\ref{k1k2}), 
the amplitude $S_{D_2}^4$ is
\be
S_{D_2}^4 \sim 16\a'^2 V^- 
\left[
\xi^+ \a'^2 \lam^2 + 
i \a' \lam \xi\cdot{\bf k}
\right],
\ee
where we used $t = 2\a' \lam = -u$.
Thus $|D^-\ra$ is not decoupled from the massless vector field.

\subsection{Summary of correlation functions}

To summarize the results of our calculation of correlation functions, 
we find that the state $|D^-\ra$ is not decoupled 
from other physical states, 
unlike all other discrete states $|D^i\ra$ and $|D^+\ra$.
The fact that the correlation functions of $|D^-\ra$ 
with an arbitrary number of tachyons always vanish 
is consistent with the possibility that 
the tachyon is a trivial representation of 
the symmetry group generated by $|D^-\ra$, 
if there is really a symmetry.
However, this fact is merely a result of 
the worldsheet time-reversal symmetry, 
and thus it might be nothing but a coincidence. 

A crucial difference between $|D^-\ra$ and other discrete states 
can also be seen from the OPE of these discrete states 
with another physical state. 
The $1/z$ term of the OPE of $|D^\mu\ra$ for $\mu\neq -$ 
with an arbitrary physical state is always again a physical state, 
since the momentum of these discrete states is zero. 
On the contrary, $|D^-\ra$ has a nonzero momentum, 
and in general its OPE with another physical state 
is no longer physical. 
By fine-tuning the momentum of the physical state in the OPE, 
one may still obtain nontrivial relations to 
constrain the correlation function, 
as it was done for all physical states in the flat background \cite{Moore}. 
But this means that we cannot define a transformation 
generated by $|D^-\ra$ on all physical states. 
This should be interpreted at most as a broken symmetry. 

On the other hand, 
the (non-)conservation law of momentum (\ref{mom-conserv})
has 26 components. 
We can match each discrete state $|D^\mu\ra$ 
with each component of this (non-)conservation law 
(with the ``anomaly'' of the conservation law matched
with the non-zero momentum of $|D^-\ra$). 
Although the $k^-$ component of (\ref{mom-conserv}) 
implies that momentum is no longer conserved, 
it also specifies precisely how it is not conserved. 
Mathematically it gives as much information 
(it imposes as much constraint on kinematics) 
as a statement of momentum conservation.
In this sense the significance of $|D^-\ra$ 
is not a bit less than any other discrete state. 
We will explore this observation further in the next section 
by studying generic field theory models with 
the same type of (non-)conservation laws.

\section{Field theory model}
\label{FieldTheory}

We would like to understand the implications of 
the (non-)conservation law (\ref{mom-conserv})
in the context of field theory. 
To begin, we consider a toy model of $N$ scalar fields 
with polynomial interactions. 
The action is of the form
\be \label{Smodel}
S = \int d^d x e^{V\cdot x} \left[
\frac{1}{2} \phi_A (- \Box + m_A^2) \phi_A + 
g_{ABC} \phi_A \phi_B \phi_C + 
\lam_{ABCD} \phi_A \phi_B \phi_C \phi_D + 
\cdots \right].
\ee
Since we have $V$ being light-like, 
the kinetic term can also be written as 
\be
\int d^d xe^{V\cdot x} \frac{1}{2} \left[ 
\del_{\mu} \phi_A \del^{\mu} \phi_A + m_A^2 \phi_A^2 
\right].
\ee

It is obvious from the Fourier decomposition of $\phi_A$ 
that the effect of the factor $e^{V\cdot X} = e^{- V^- X^+}$ 
to the interaction vertices is simply to modify 
the momentum conservation law to 
\be \label{vertex}
\sum_{a=1}^n k_a + iV = 0
\ee
for an $n$-point vertex with incoming momenta $k_a$. 
This is in agreement with the (non-)conservation law 
(\ref{mom-conserv}) for the linear dilaton background
for $\chi_{\cal M} = 1$. 

The propagator in Fourier basis is given by
\be
G(k, k') = \frac{1}{k\cdot k' - m^2} \delta(k + k' - iV).
\ee

For a generic Feynman diagram with 
$v$ vertices, $e$ propagators and $f$ loops
\footnote{
Here we draw the diagram on a closed surface 
so that the number of loops is always one more 
than the value one usually counts in field theory 
due to the ``outer loop'' which closes the surface. 
}
(which is a diagram with $v$ vertices, $e$ edges 
and $f$ faces), 
we can draw the diagram without line crossing 
on a closed surface $C$ with Euler character 
\be
v - e + f  = \chi_C \equiv 2 - 2g' - c',
\ee
where $g'$ is the number of handles and $c'$ the number of cross-caps.
The action (\ref{Smodel}) tells us to associate 
a factor of $e^{V\cdot X}$ for each vertex, 
and a factor of $e^{-V\cdot X}$ for each propagator,
so that the momenta of the external legs of a Feynman diagram 
satisfy the relation 
\be \label{mom-conserv2}
\sum_a k_a + i (v-e) V = 0.
\ee
The coefficient $(v-e)$ here 
should be compared with its counterpart $\chi_{\cal M}$ in 
the (non-)conservation law (\ref{mom-conserv}) 
in the linear dilaton background
\be
v-e = \chi_C - f = 2 - 2g' - f - c', \qquad
\chi_{\cal M} = 2 - 2g - b - c.
\ee
Imagining that $\phi_A$'s represent spacetime fields 
of open string oscillation modes, 
and that a Feynman diagram with propagating lines 
replaced by strips is the string worldsheet, 
we would identify a loop in the Feynman diagram 
as a boundary of the worldsheet Riemann surface, 
and therefore we are led to the identification
\be
b = f, \qquad g = g', \qquad c = c'.
\ee
Thus the two (non-)conservation laws are exactly the same! 
We believe that this toy model captures the main features 
of the (non-)conservation law (\ref{mom-conserv}) 
in the linear dilaton background.

Due to its close relationship with string theory
in light-like linear dilaton background, quantum
field theory models with actions of the form (\ref{Smodel})
deserve further discussion. 

For some special cases, for example, if 
\be
S = \int d^d x e^{V\cdot x} \left[
\frac{1}{2} \phi_A \Box \phi_A + 
g_{ABC} \phi_A \phi_B \phi_C 
\right],
\ee
it is possible to define a symmetry transformation 
\bea
&x^i \rightarrow L x^i, \qquad x^- \rightarrow L^2 x^-, 
\qquad x^+ \rightarrow x^+ + a, \\
&\phi_A \rightarrow L^{-2}\phi_A, \qquad 
L \equiv e^{-\frac{V\cdot a}{d-6}}, 
\eea
such that the action $S$ is invariant. 
However, for generic interaction terms and masses, 
it seems impossible to define symmetry transformation 
under which the action (\ref{Smodel}) is invariant. 

Recall that the translation symmetry 
$x^{\mu} \rightarrow x^{\mu} + a^{\mu}$ 
is fully encoded in the constraint of momentum conservation 
\be \label{ka=0}
\sum_a k_a = 0, 
\ee
and is equivalent to the requirement that
the action is of the form 
\be \label{flat}
S = \int d^d x\; {\cal L}(\phi, \del \phi, \cdots), 
\ee
where ${\cal L}$ has no explicit dependence on $x^{\mu}$. 
Similarly, the (non-)conservation law (\ref{mom-conserv2})
implies that the action is of the form 
\be \label{eVx}
S = \int d^d x\; e^{V\cdot x}{\cal L}(\phi, \del \phi, \cdots).
\ee
Since the implication of 
either momentum conservation (\ref{ka=0}) or 
(non-)conservation (\ref{mom-conserv2}) 
are equally powerful constraints, 
it is a little odd that only one of them always comes 
from a symmetry. 
This should be taken as a hint 
that our notion of symmetry 
should be generalized to accommodate 
the momentum (non-)conservation law 
and other similar cases.

It is tempting to make the conjecture that 
the string field theory action in light-like linear dilaton background 
is related to that in flat spacetime 
in the same way (\ref{eVx}) is related to (\ref{flat}), 
probably up to certain field redefinitions and gauge fixing. 
It will be very interesting to check this explicitly.

Incidentally, we remark that 
the factor $e^{V\cdot X}$ 
of the kinetic term can be removed by a field redefinition
\be
\phi_A = e^{-\frac{1}{2} V \cdot X} \psi_A.
\ee
Since
\be
e^{V\cdot X} \del_{\mu}\phi_A \del^{\mu}\phi_A
= \del_{\mu}\psi_A \del^{\mu}\psi_A - \frac{V^-}{2}\del^+(\psi_A^2),
\ee
the action (\ref{Smodel}) becomes
\be 
S = \int d^d x \left[
\frac{1}{2} \psi_A (\Box + m_A^2) \psi_A + 
g_{ABC} e^{-V\cdot X/2} \psi_A \psi_B \psi_C + 
\lam_{ABCD} e^{-V\cdot X} \psi_A \psi_B \psi_C \psi_D + 
\cdots \right].
\ee
The kinetic term is now canonical, 
and the $n$-point vertex receives a factor of 
$e^{(n-2)V\cdot X/2}$,
giving the new modified conservation law
\be
\sum_{a=1}^n k_a - i\frac{(n-2)}{2}V = 0.
\ee
This is of course just (\ref{vertex}) with $k_a \rightarrow k_a - iV/2$.

An interesting property of this class of models (\ref{Smodel}) 
is that due to the difference in (non-)conservation laws (\ref{mom-conserv2})
for different topologies of Feynman diagrams, 
the contribution of quantum corrections 
is separated from the classical tree level amplitude. 
For example, the correlation function
\be
\la \tilde{\phi}_A(k_1) \tilde{\phi}_B(k_2) \tilde{\phi}_C(k_3) \ra
\ee
receives no quantum correction at all if 
$k_1 + k_2 + k_3 + i V = 0$.

\section{Concluding remarks} \label{Comments}

In this paper we propose that physical states 
which we call ``discrete states'' play special roles in string theory. 
It is well known that they generate a huge symmetry in 2D. 
They also generate the translation symmetry 
in 26D flat spacetime.
We found that there are also discrete states 
in the 26D light-like linear dilaton background. 
Interestingly, in addition to the discrete states
corresponding to the translation symmetry in the transverse directions
($|D^i\ra$, $|D^+\ra$), 
we have a discrete state $|D^-\ra$ 
which does not seem to have a simple interpretation
as a symmetry generator.

There are a few facts to keep in mind. 
First, the discrete state $|D^-\ra$ has zero norm. 
Usually, zero-norm states are also spurious physical states.
$|D^-\ra$ is an exception to this general rule.
We do not demand discrete states 
to obey the usual rule that zero-norm states also have to be spurious, 
or that negative-norm states have to be decoupled.
Usually, zero-norm and negative-norm states 
imply problems with unitarity if they are not decoupled.
If they are coupled to other physical states,
there can be a non-zero probability 
to create zero-norm or negative-norm state from 
the scattering of physical states, 
and unitarity is broken.
However, if the zero-norm or negative-norm state is a discrete state, 
the volume of the phase space available in a scattering process is zero, 
so the probability of creating such states in a physical process is zero,
and unitarity is not broken.
Therefore, unitarity does not imply that 
zero-norm and negative-norm
discrete states must decouple from physical states.

Having said this, 
we note that all discrete states except $|D^-\ra$
are decoupled from all physical states, 
even though they are not spurious states.
This may seem a little puzzling at first sight.
However this is just a result of the fact that 
all spacetime fields are in the trivial representation 
of the translation group.
The fact that $|D^-\ra$ is a state with zero norm, 
and the fact that it is algebraically analogous to 
the discrete states in 2 dimensional string theory, 
still strongly suggest that
it should play a special role in the theory.

The clue of the role played by $|D^-\ra$ lies in the observation that,
although the translation symmetry in the $X^+$ direction 
is broken by the dilaton background,
we still have $26$ (non-)conservation laws (\ref{mom-conserv})
for the external momenta, 
just like in flat space. 
The (non-)conservation law is equally powerful 
in constraining the dynamics of the theory 
for an arbitrary value of $V$, including $V = 0$. 
In this sense the ``symmetry'' 
of the linear dilaton background
is as big as the flat spacetime.
The vertex operator $|D^-\rangle$ corresponding 
to the momentum non-conservation in the $k^-$ direction
\be
\sum_a k_a^- + i\chi_{\cal M} V^- = 0
\ee
is therefore playing the same role
as all other discrete states $|D^i\rangle$, and $|D^+\rangle$,
which correspond to the conservation law
\be
\sum_a k_a^i = 0, \qquad
\sum_a k_a^+ = 0.
\ee

String theory is known to have huge hidden symmetries 
which ensure all the nice properties 
such as dualities and self-consistency. 
A clear and explicit understanding of these symmetries 
is however never in reach except for 2D strings. 
Perhaps this is because our concept of symmetry 
is still too primitive. 
To conclude, we believe it is important
to study field theory models 
with constraints mimicking the effect of symmetries, 
and then try to generalize the notion of symmetry to 
incorporate these structures.

\section*{Acknowledgements}

We thank Chong-Sun Chu, Kazuyuki Furuuchi, 
Takeo Inami, Hsien-Chung Kao, Yu Nakayama,
Shunsuke Teraguchi, Wen-Yu Wen and Syoji Zeze 
for valuable discussions.
The work is supported in part by
the National Science Council, 
and the National Center for Theoretical Sciences, Taiwan, R.O.C.

\appendix

\section*{Appendix: path integral for the linear dilaton background}

In this appendix we show how to compute the correlation function in 
the light-like linear dilaton background. 
A linear dilaton background modifies the correlation function in two ways. 
First, it changes the worldsheet boundary condition in the open string case. 
Second, it modifies the momentum conservation law.
The worldsheet action for an open string in 
the linear dilaton background is given by
\be
S=\frac1{4\pi\a'}\int_{\cal M}d^2\s\,\sqrt{g}
g^{ab}\del_a X\cdot \del_b X
+\frac1{4\pi}\int_{\cal M}d^2\s\,\sqrt{g} R(\s)V\cdot X(\s)
+\frac1{2\pi}\int_{\del \cal M}ds \, \k(\s)V\cdot X(\s),
\ee
and its variation is
\be 
\d S=-\frac 1{2\pi\a'}\int_{\cal M}d^2\s\,\sqrt{g}
(\Lap X-\frac {\a'}2 R(\s)V)\cdot \d X-\frac1{2\pi\a'}
\int_{\del \cal M}ds\,(\del_nX-\a'\k(\s)V)\cdot \d X,
\ee
where $R(\s)$ ($\k(\s)$) is the worldsheet curvature
(geodesic curvature of the boundary), 
and $\del_n$ denotes the normal derivative along the boundary. 
From this expression, we can read off the equation of motion 
\be
\Lap X^\m(\s)=\frac{\a'}2R(\s)V^\m, \qquad \s \in {\cal M},
\ee
and boundary condition
\be
\del_n X^\m(\s)=\a'\k(\s)V^\m, \qquad \s \in \del{\cal M}
\ee
for an open string.
The Stoke's theorem implies
\be
\int_{\del \cal M}ds\, \del_nX^\m=\int_{\cal M}d^2\s\,
\sqrt{g}\Lap X^\m,
\ee
so that
\be \label{a1}
\int_{\del \cal M}ds\,\k(\s)=\frac12\int_{\cal M} 
d^2\s\,\sqrt{g}R(\s).
\ee
On the other hand, the Gauss-Bonnet theorem says
\be \label{a2}
\frac1{2\pi}\int_{\del \cal M}ds\,\k(\s)+
\frac1{4\pi}\int_{\cal M}d^2\s\,\sqrt{g} R(\s)=\chi_{\cal M},
\ee
where $\chi_{\cal M}$ is the Euler character of 
the worldsheet (\ref{Euler}). 
These two equations (\ref{a1}, \ref{a2}) give 
\be 
\frac1{2\pi}\int_{\del \cal M}ds\,
\k(\s)=\frac1{4\pi}\int_{\cal M}d^2\s \sqrt{g}\,R(\s)
=\frac {\chi_{\cal M}}2.
\ee
Using worldsheet conformal symmetry, 
one can set both $R(\s)$ and $\k(\s)$ constant
\be 
\k(\s)={\pi{\chi}_{\cal M} \over \Length}, 
\qquad
R(\s)= {2\pi{\chi}_{\cal M} \over \Area},
\ee
where $\Length$ and 
$\Area$ are the boundary length and worldsheet area, respectively.

Separating the solution $X^\m(\s)$ into a special solution and 
a homogeneous solution, 
we find the equation of motion and boundary condition simplified as
\begin{align} 
&\Lap X^\m_s(\s) = {\a'\pi\over \Area}{\chi}_{\cal M}V^\m,  \quad
&\Lap X^\m_h(\s)  = 0, \qquad \qquad
&(\s \in {\cal M}), \\
&\del_n X^\m_s(\s) = {\a'\pi\over \Length}{\chi}_{\cal M}V^\m, \quad
&\del_nX^\m_h(\s) = 0, \qquad \qquad
&(\s \in \del{\cal M}). 
\end{align}

The homogeneous solution satisfies exactly the same set of equations 
as in the flat space. 
Therefore, the Green's function in linear dilaton background is
\be
G'(\s_1,\s_2)=-\frac {\a'}2 \ln\left|z_1-z_2\right|^2-\frac {\a'}2 \ln\left|z_1-\bar z_2\right|^2+f(z_1,\bar z_1)+f(z_2, \bar z_2),
\ee   
where 
\be
f(z,\bar z)={\a'{\chi}_{\cal M}V^\m \over 8\Area}
\int_{\cal M}d^2z'\,e^{2\om (z',\bar z')}\ln\left|z-z'\right|^2 
+ \mbox{constant}
\ee
comes from the special solution, 
and the constant is determined by requiring that 
$G'$ is orthogonal to the zero mode of $X^\m(\s)$.

Now we are ready to derive the correlation function 
in the linear dilaton background. 
The generating functional is
\begin{align}
Z[J]&=\left\la \exp\left(i\int d^2\s\, J(\s)\cdot X(\s)\right)\right\ra \nn \\
&=\int DX \; \exp\left\{\int_{\cal M} d^2\s\,\left[\sqrt{g}
\left(X^\m\frac\Lap{4\pi\a'} X_\m-\frac{R(\s)}{4\pi}V\cdot X\right)
+iJ(\s)\cdot X(\s)\right]\right\} \nn \\
& \quad \times \exp\left\{\frac1{4\pi\a'}\int_{\del\cal M}ds\, X \cdot
\left[\del_nX-2\a'\k(\s)V\right]\right\}. 
\end{align}
Expanding $X^\m(\s)$ in terms of a complete basis $X_I(\s)$, 
we have
\begin{align}
X^\m(\s)=\sum_Ix_I^\m X_I(\s), \qquad
\Lap X_I=-\om_I^2X_I, \qquad
\left.\del_nX_I\right|_{\del\cal M}=0, \nn\\
\int_{\cal M}\d^2\s\,g^{1/2}X_IX_J=\d_{IJ},\qquad 
X_0 =\left(\int_{\cal M}d^2\s\,g^{1/2}\right)^{-1/2}
= \frac1{\sqrt{\Area}}.
\end{align}
Then
\begin{align}
-\frac1{4\pi}\int_{\cal M}d^2\s\,\sqrt{g} R(\s)V\cdot X(\s)&
=-\frac{{\chi}_{\cal M}}{2\sqrt{\Area}}V_\m x_0^\m
-\sum_{I\neq0}V_\m x_I^\m R_I, \nn\\
-\frac1{2\pi}\int_{\del\cal M}ds\,\k(\s)V\cdot X(\s)&
=-\frac{{\chi}_{\cal M}}{2\sqrt{\Area}}V_\m x_0^\m
-\sum_{I\neq0}V_\m x_I^\m\k_I, \nn\\
i\int_{\cal M}d^2\s\,J(\s)\cdot X(\s)&
=iJ_0\cdot x_0+i\sum_{I\neq0}J_I\cdot x_I,  
\end{align}
where
\begin{align}
R_I&=\int_{\cal M}d^2\s\,\sqrt{g}{R(\s)\over4\pi}X_I(\s), \nn\\
\k_I&=\int_{\del\cal M}ds\,{\k(\s)\over 2\pi}X_I(\s), \nn\\
J_I^\m&=\int_{\cal M}d^2\s\,J^\m(\s)X_I(\s).
\end{align}

Using the worldsheet conformal symmetry, 
we have distributed the curvature uniformly on the world sheet,
i.e. $R(\s)=R=\mbox{constant}$ 
and $\k(\s)=\k=\mbox{constant}$. 
This greatly simplifies the calculation 
since now we have $R_I=\k_I=0$ from 
the orthogonality condition of $X_I(\s)$. 
It is straightforward to evaluate the zero mode integral as
\be
\prod^{d-1}_{\m=0}\int dx_0^\m \;
\exp\left\{i\left(i{{\chi}_{\cal M}\over \sqrt{\Area}}V+J_0\right)\cdot x_0\right\} 
=i(2\pi)^d \Area^{d/2}\d^d\left(\sqrt{\Area}J_0+i{\chi}_{\cal M}V\right),
\ee
where the factor $i$ comes from the Wick rotation 
$x_0^0\rightarrow -ix_0^d$. 
      
The non-zero mode integral is easy to handle as well. 
We have
\begin{align}
Z[J]&=(\mbox{zero\;mode\;part})\times
\prod_{I\neq0,\m}\int dx_I^\m\,
\exp\left(-{\om_I^2x_I^\m x_{I\m}\over 4\pi\a'}
+ix_I^\m J_{I\m} +\frac1{4\pi\a'}\int_{\del\cal M}ds\,
x_I^\m x_{J\m}X_I\del_n X_J\right) \nn\\
&=i(2\pi)^d \Area^{d/2}\d^d
\left(\sqrt{\Area}J_0+i{\chi}_{\cal M}V\right)
\left({\det}'{-\Lap\over 4\pi^2\a'}\right)^{-d/2} 
e^{-\frac12\int_{\cal M} d^2\s d^2\s'
\,J(\s)\cdot J(\s')G'(\s,\s')}, 
\end{align}
where in the second line 
we have used the boundary condition 
$\left.\del_nX_I\right|_{\del\cal M}=0$
to eliminate the boundary term.
The notation $\det'$ means the determinant 
defined without the zero modes. 

Consider the path integral with a product of tachyon vertex operators
\be 
A_{\cal M}^n(k,\s)=\left\la e^{ik_1\cdot X(\s_1)} e^{ik_2\cdot X(\s_2)}
\cdots e^{ik_n\cdot X(\s_n)} \right\ra_{\cal M}. 
\label{ampl}
\ee
This corresponds to 
$$
J^\m(\s)=\sum^{n}_{a=1}\,k_a^\m\d^2(\s-\s_i).
$$
In particular, 
$J_0^\m=\frac1{\sqrt {\Area}}\left[\sum^{n}_{a=1}k_a^\m\right]$.

The amplitude (\ref{ampl}) then becomes
\begin{align}
A_{\cal M}^n&=iC_{\cal M}^X(2\pi)^d\d^d\left(
{\sum}_a k_a^\m+i{\chi}_{\cal M}V^\m\right)\exp
\left(-\frac12\int_{\cal M} d^2\s d^2\s'\,
J(\s)\cdot J(\s')G'(\s,\s')\right),
\end{align}
where
\be
C_{\cal M}^X=\Area^{d/2}
\left({\det}'\frac{-\Lap}{4\pi^2\a'}\right)_{\cal M}^{-d/2}
=\mbox{constant}.
\ee

More generally,
\begin{align}
\left\la \prod^{n}_{a=1} e^{ik_a\cdot X(y_i)}\, 
\prod^{p}_{b=1}\del_yX^{\m_b}(y_b') \right\ra_{D_2}
&= iC_{\cal M}^X(2\pi)^d\d^d
\left({\sum}_a k_a^\m+i{\chi}_{\cal M}V^\m\right) \nn\\
&\times \prod^{n}_{a,b=1 a<b}\left|y_a-y_b\right|^{2\a'k_a\cdot k_b} 
\left\la\prod^{p}_{b=1} [v^{\m_b}(y_b')+q^{\m_b}(y_b')]\right\ra_{D_2},
\end{align}
where $v^\m(y)=-2i\a'\sum^{n}_{a=1}\frac{k_a^\m}{y-y_b}$, 
and $q$'s are contracted using $-2\a'(y-y')^{-2}\eta^{\m\n}$.

Since the linear dilaton background does not affect the ghost action, 
the calculation for ghost contribution remains the same. 
We can still fix three open string vertex operators 
on the boundary and compensate it with the corresponding ghost determinant.

\end{document}